\documentclass[twocolumn,preprintnumbers,amsmath,amssymb]{revtex4}
\usepackage{tipa}
\usepackage{amssymb}
\usepackage{stmaryrd}
\usepackage{txfonts}

\usepackage{graphicx}
\usepackage{dcolumn}
\usepackage{bm}

\begin{document}

\preprint{}

\title{Precision Measurement of the Optical Conductivity of Atomically Thin Crystals via Photonic Spin Hall Effect}
\author{Shizhen Chen}
\author{Xiaohui Ling}
\author{Weixing Shu}
\author{Hailu Luo}\email{hailuluo@hnu.edu.cn}
\author{Shuangchun Wen}
\affiliation{Laboratory for Spin Photonics, School of Physics
and Electronics, Hunan University, Changsha 410082, China}

\date{\today}

\begin{abstract}
How to measure the optical conductivity of atomically thin crystals is an important but challenging issue due to the weak light-matter interaction at the atomic scale. Photonic spin Hall effect, as a fundamental physical effect in light-matter interaction, is extremely sensitive to the optical conductivity of atomically thin crystals. Here, we report a precision measurement of the optical conductivity of graphene, where the photonic spin Hall effect acts as a measurement pointer. By incorporating with the weak-value amplification
technique, the optical conductivity of monolayer graphene taken as a universal constant of $(0.993\pm0.005)\sigma_0$ is detected, and a high measuring resolution with $1.5\times10^{-8}\Omega^{-1}$ is obtained. For few-layer graphene without twist, we find that the conductivities increase linearly with layer number. Our idea could provide an important measurement
technique for probing other parameters of atomically thin crystals, such as magneto-optical constant, circular dichroism, and
optical nonlinear coefficient.
\end{abstract}


\maketitle
Atomically thin crystals have extraordinary
electronic and photonic properties which hold great promise
in the applications of photonics and optoelectronics~\cite{Bonaccorso2010}.
A characterization of the parameters of the atomically thin crystals is therefore essential for
photonics and optoelectronics applications.
Graphene, a several atomic layer of graphite, has attracted much attention due to its novel optical and electrical properties~\cite{Novoselov2004}. The optical properties of graphene, such as the remarkable absorption, reflectivity, and plasmonic response, are closely related to its optical conductivity~\cite{Gusynin2006,Falkovsky2008,Low2014}.
In general, optical measurement extracts information from the measured system via light-matter interaction.
Therefore, how to measure the optical conductivity of atomically thin crystals is still a challenging issue due to the weak light-matter interaction at the atomic scale.

There exist some confusion about the actual value of the optical conductivity. The theoretical prediction has shown that the conductivity generally is a frequency-dependent value~\cite{Falkovsky2007}. At low photon energies, the conductivity is contributed from intraband transitions. However, it is argued that in the hight-frequency range the optical conductivity should be a universal constant $\sigma_0=e^2/4\hbar$ due to interband processes~\cite{Ziegler2007,Kuzmenko2008,Stauber2008}. The optical spectroscopy measurements also yield
a universal dynamic conductivity $(1.01\pm0.04)\sigma_0$ over the visible frequency range~\cite{Nair2008}. This approach can measure the optical conductivity with optical absorption in a wide spectral regime~\cite{Dawlaty2008,Mak2008}. While at lower photon energies a departure from the universal constant has been observed. But another experimental measurement shows that the conductivity rises smoothly and steadily in the visible spectral range~\cite{Mak2011}. Therefore, it still remains ambiguous whether the universal hypothesis holds up for the optical conductivity at visible frequencies. Moreover, due to the weak interaction between graphene and light, an advanced measurement method and a high measurement resolution are urgently needed.

\begin{figure}
\includegraphics[width=8cm]{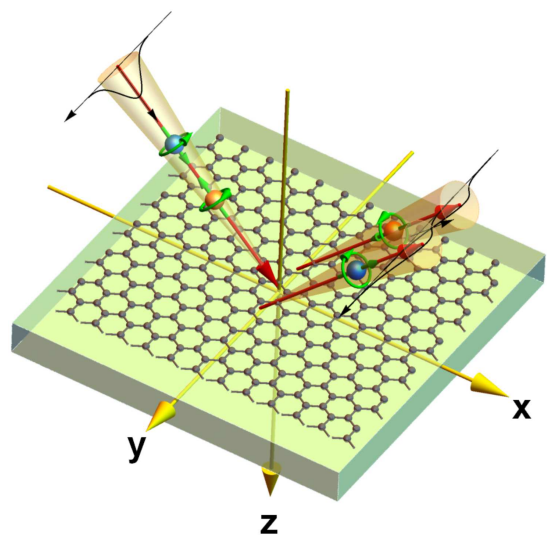}
\caption{\label{Fig1} Illustration of photonic spin Hall effect at a air-graphene-medium interface. The reflecting light beam experiences a spin dependent shift due to the spin-orbit coupling on the surface of graphene.}
\end{figure}

In this Letter, photonic spin Hall effect (SHE)~\cite{Onoda2004,Bliokh2006,Yin2013,Shitrit2013}, as a fundamental physical effect in light-matter interaction, is proposed to realize the precision measurement of the optical conductivity of atomically thin crystals. The photonic SHE, including its underlying physics and possible applications, has been extensively studied in recent years~\cite{Bliokh2015,Ling2017}. This effect is attributed to the spin-orbit coupling during the process of light-matter interaction. Photonic SHE manifested itself as a spin-dependent shift provides important information of optical interface~\cite{Zhou2012I}, and therefore it can be employed as the pointer in precision measurements. In fact, photonic SHE is a weak effect whose initial shift is only a fraction
of a wavelength, and therefore can not be detected directly by conventional optical measurements. By incorporating with the weak-value amplification techniques~\cite{Aharonov1988}, however, the initial shift can be enhanced by
nearly four orders of magnitude~\cite{Hosten2008}. The observation of the photonic SHE with a sensitivity of $1\AA$ have been achieved based on the quantum weak measurements.

A general model treats atomically thin crystal as a homogeneous slab with an effective thickness and a certain refractive index~\cite{Blake2007,Bruna2009,Zhou2012II,Chen2017}. However, as a truly atomic crystal whose thickness is much less than the wavelength of light, the refractive index seems not well defined. It has been shown that the model with zero-thickness can more accurately describe the optical response of graphene~\cite{Merano2016}. This zero-thickness model is well established by considering the atomically thin crystals as a two-dimensional conductive film. Let us consider that an electromagnetic wave propagating in a dielectric medium with refractive index $n_1$ impinges on the flat interface with
a second medium with refractive index $n_2$, which is coated by few-layer graphene.
In the two media, the electric (magnetic) fields are respectively represented by $\mathbf{E}_{1}$ and $\mathbf{E}_{2}$ ($\mathbf{H}_{1}$ and $\mathbf{H}_{2}$). $\mathbf{E}$ is related to $\mathbf{H}$ in vacuum by the impedance $Z_0$ of vacuum. Taking the surface current density into account, the boundary conditions are $\hat{n}\times(\mathbf{E}_{2}-\mathbf{E}_{1})=0$, $\hat{n}\times(\mathbf{H}_{2}-\mathbf{H}_{1})=\mathbf{J}_{s}+\mathbf{J}_{p}$. $\hat{n}=-\hat{z}$ is the unit vector normal to the interface. $\mathbf{J}_{s}=\sigma\mathbf{E}$ is from the Ohm's law, where $\sigma$ is the optical conductivity of the atomically thin crystal.  $\mathbf{J}_{p}=\partial \mathbf{P}/\partial t$ is a current density depends on the change of polarization $\mathbf{P}=\varepsilon_0 \chi \mathbf{E}$ with $\varepsilon_0$ and $\chi$ being permittivity and electric susceptibility, respectively.
Based on the boundary condition, the Fresnel coefficients can be obtained as~\cite{Kamp2015,Cai2017}
\begin{equation}
r_{p}=\frac{n_2\cos\theta_i-n_1\cos\theta_t+(ik\chi+\sigma Z_0)\cos\theta_i\cos\theta_t}{n_2\cos\theta_i+n_1\cos\theta_t+(ik\chi+\sigma Z_0)\cos\theta_i\cos\theta_t}\label{RPP},
\end{equation}
\begin{equation}
r_{s}=\frac{n_1\cos\theta_i-n_2\cos\theta_t-ik\chi-\sigma Z_0}{n_1\cos\theta_i+n_2\cos\theta_t+ik\chi+\sigma Z_0}\label{RSS},
\end{equation}
where $\theta_i$ and $\theta_t$ are the angles of incidence and refraction, respectively (see the Supplemental Material). If $\sigma$ and $\chi$ are both zero, the Fresnel coefficients reduce to the ordinary forms at air-glass interface.

\begin{figure}
\includegraphics[width=8cm]{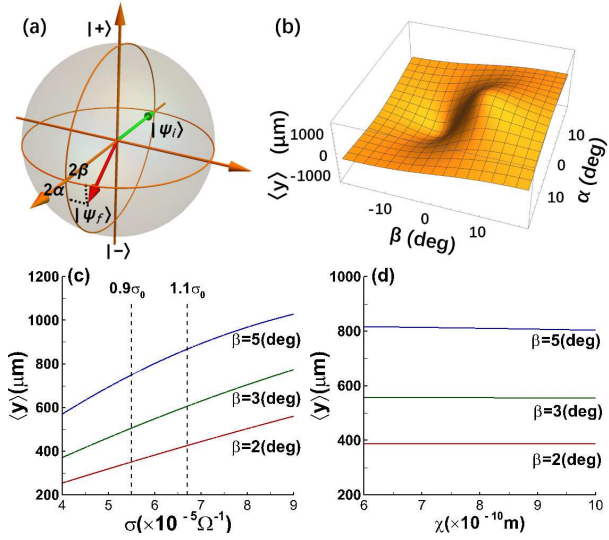}
\caption{\label{Fig2} Amplification of the photonic spin Hall effect with preselected and postselected ensembles. (a) Representation of the initial and final states on the Poincar\'{e} sphere. (b) Amplification of spin-dependent splitting as a function of the angles $\alpha$ and $\beta$. The amplified shifts vary with (c) optical conductivity and (d) electric susceptibility. The dashed line indicates the conductivity value corresponding to $0.9\sigma_{0}$ and $1.1\sigma_{0}$. }
\end{figure}

 When a light beam imping on the surface of graphene, it experiences photonic SHE manifested itself as a tiny separation of two spin components, as shown in Fig.~\ref{Fig1}. Here, the Gaussian beam can be regarded as the bounded beam which can be regarded as the combination of plane electromagnetic waves. As the photonic SHE appears in $y$ direction, we only consider wavevector ${k}_{y}$ component in the momentum space. Supposing the polarization of the beam is horizontal polarization, the polarization states of different angular spectrum components can be written as $|{{H}}({k}_{y})\rangle$. Each angular spectrum components is related to the reflection coefficients dependent on its component of wavevector. After reflecting at the graphene interface, the rotations of polarizations for each angular spectrum components are different to satisfy the photon transversality. Under the paraxial approximation, the change in the polarization state after reflection becomes
\begin{equation}
 |{H}({k}_{y})\rangle\rightarrow r_{p}|{H}({k}_{y})\rangle-\frac{k_{y}\cot\theta_i(r_{p}+r_{s})}{k_{0}}|{V}({k}_{y})\rangle\label{HKI}
\end{equation}
Here, $k_0=\omega/c$ is the wavevector in vacuum. $|{{V}}({k}_{y})\rangle$ is vertical polarization caused by tiny rotation of the polarization, which originates from the spin-orbit coupling of light.

To interpret the spin-orbit coupling more clearly, the eigenstates can be written in the spin basis by the relations $|{H}\rangle=\frac{1}{\sqrt{2}}(|{+}\rangle+|{-}\rangle)$ and $|{V}\rangle=\frac{1}{\sqrt{2}}i(|{-}\rangle-|{+}\rangle)$. Under the condition of weak interaction $k_{y}\delta\ll1$, Eq.~(\ref{HKI}) becomes
\begin{equation}
|{H}({k}_{y})\rangle\rightarrow \exp(-k_{y}\delta)|{+}\rangle+\exp(k_{y}\delta)|{-}\rangle
\end{equation}
where $\delta$ represents the spin-dependent shift induced by the spin-orbit interaction, which is given by
\begin{equation}
 \delta=[(r_{p}+r_{s})\cot\theta_i]/(k_0r_{p}).\label{H}
\end{equation}
Note that the spin-dependent shift is determined by the Fresnel coefficients, which is related to the optical conductivity (see the Supplemental Material).
Therefore, the photonic SHE can serve as a pointer in the measurements of the optical conductivity of atomically thin crystals.
The wavefunction in the spin basis $|s\rangle$ ($s=\pm1$) can be written as $|{\psi}\rangle=\int\exp(-k_{y}\hat{\sigma_3}\delta)|\Phi(k_{y})\rangle|s\rangle dk_{y}$ with $|\Phi(k_{y})\rangle$ being a Gaussian distribution. Here, $\exp(-k_{y}\hat{\sigma_3}\delta)$ indicates the spin-orbit coupling of light with $\hat{\sigma_3}$ being the Pauli operator.  Therefore, the optical conductivity of graphene can be precisely measured by incorporating with the weak-value amplification techniques.

\begin{figure}
\includegraphics[width=8cm]{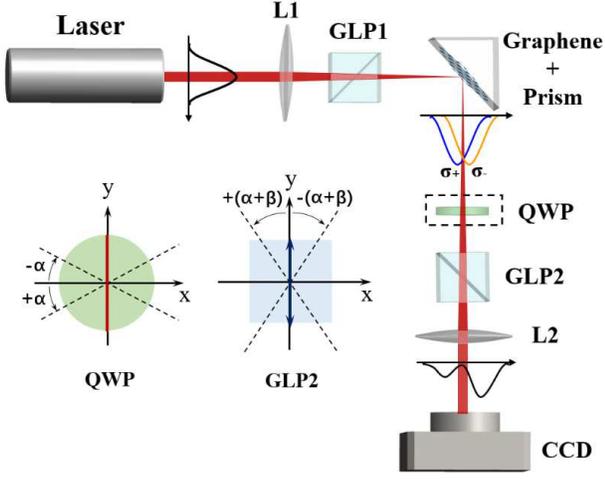}
\caption{\label{Fig3} Schematic of the experimental setup to detect the tiny spin-orbit interaction of light in graphene. A polarized Gaussian beam generated by He-Ne laser is incident at a angle on the graphene interface. The lenses of L1 and L2 can focus and collimate the light beam. The Glan laser polarizers (GLP1
and GLP2) select the preselected and postselected states, respectively. The CCD is used for capturing the intensity
profiles.}
\end{figure}

In a quantum weak measurement, a observable of a system is first coupled to a separate degree of freedom, and then the information about the state of the observable is read out from the meter~\cite{Kofman2012,Dressel2014}.
Here, the observable is the spin of photon and the spin-dependent shifts serve as the pointer of meter. In a weak measurement scheme, the coupling between the observable and the meter is weak, and only a small amount of information is extracted. In the linear theory with approximation, the output result of the weak measurement is a extraordinary value called weak value. Weak value is naturally determined by the preselection state $|\psi_{i}\rangle$ and postselection state $|\psi_{f}\rangle$, which is given by
\begin{equation}
A_{w}=\frac{\langle{\psi_{f}}|\hat{\sigma_3}|{\psi_{i}}\rangle}{\langle{\psi_{f}}|{\psi_{i}}\rangle},
\end{equation}
where $\hat{\sigma_3}$ is the operator of an observable (see the Supplemental Material). A large weak-value can be obtained when the preselection state and postselection are close to orthogonal $\langle{\psi_f}|{\psi_i}\rangle\approx0$.

 We choose the preselected state as $|{H}\rangle=(|+\rangle+|-\rangle)/\sqrt{2}$ in the weak measurement. To achieve a large weak value, the postselection state $|\psi_{f}\rangle$ needs to be nearly orthogonal to $|\psi_{i}\rangle$. Then the postselection state is chosen as $|\psi_{f}\rangle=\sin\left(\pi/4+\alpha\right)|+\rangle-e^{2i\beta}\cos\left(\pi/4+\alpha\right)|-\rangle$, with
$\alpha$ and $\beta$ being the small deviation angles as shown on the Poincar\'{e} sphere [Fig.~\ref{Fig2}(a)]. From the above preselection and postselection states, the weak value $A_w$ of the observable $\hat{\sigma_3}$ is a complex number~\cite{Xu2013,Jordan2014}. The wavefunction after the preselection and postselection ensembles becomes $\langle\psi_{f}|{\psi}\rangle=\int\exp(-k_{y}A_{w}\delta)|\Phi(k_{y}) dk_{y}$ (see the Supplemental Material). The real and imaginary parts of the weak value is related to the angles $\alpha$ and $\beta$, which amplifies the spin-dependent shifts as shown in Fig.~\ref{Fig2}(b). The spin-dependent shifts in photonic SHE serve as a pointer, the imaginary part of the weak value would significant amplification due to the free evolution of the wave function. To effective detect the photonic SHE, a purely imaginary weak value is given by $A_w=i\cot\beta$, with $\alpha=0$ in our experiment.

In our case, a free evolution factor $F$ related to the meter state for the wavefunction evolution in real position space~\cite{Aiello2008}. The factor given by $F=z/z_r$ amplifies the initial pointer shift together with the weak value, where $z$ is the effective evolution distance. In the linear theory of weak measurements~\cite{Aharonov1988}, the final amplified pointer shift is given by
\begin{equation}
\langle{y}\rangle= F \mathrm{Im}[A_{w}]\delta=\frac{z}{z_r}\cot{\beta}\delta\label{AS}.
\end{equation}
From Eq.~(\ref{AS}), we can achieve an amplified pointer shift by an amount about $F\cot\beta$ times larger than the initial pointer shift via weak measurements. The above equation is the simplest form of the shift, which can explain the underlying physics of weak measurements clearly. However, it would fail to describe the weak-value amplification for small preselected angle $\beta$. The strict expression of the amplified pointer shift can be obtained as:
\begin{equation}
\langle{y}\rangle=\frac{zr_p(r_p^{'}+r_s^{'})\cot\theta_i\sin2\beta}{(r_p^{'}+r_s^{'})^2\cos^2\beta\cot^2\theta_i+[2k_0 z_r(r_p^{'2}+r_p^{''2})+\xi^2]\sin^2\beta}\label{Ap},
\end{equation}
where $\xi=\mathrm{Re}[\partial r_p/\partial\theta_i]$ from the Taylor series expansion, the complex reflectance coefficients are defined as $r_{p,s}=r_{p,s}^{'}+ir_{p,s}^{''}$. The detailed calculation can been seen in the Supplemental Material.

\begin{figure}
\includegraphics[width=8cm]{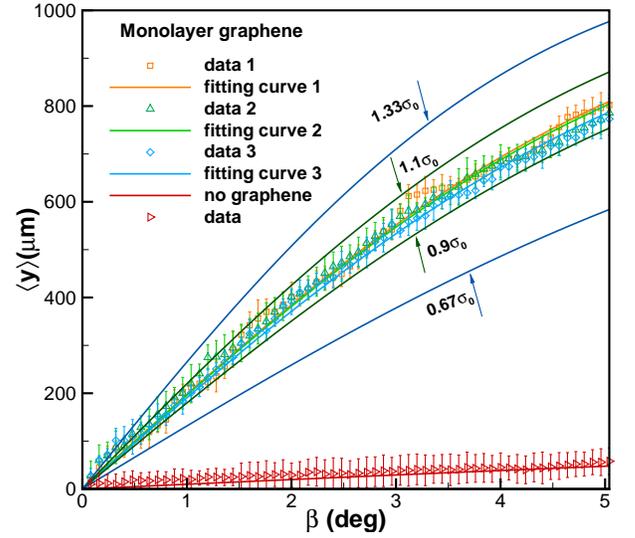}
\caption{\label{Fig4} Amplification pointer shifts as a function of the postselection angle $\beta$. Theoretical results for optical conductivity taken as $0.67\sigma_0$, $0.9\sigma_0$, $1.1\sigma_0$, and $1.33\sigma_0$. The fitting curves are obtained based on three groups of data. For comparison, the result without graphene is also given.}
\end{figure}

To demonstrate sensitivity of the photonic SHE to the optical conductivity, the calculation from Eq.~(\ref{Ap}) with the change of the conductivity of monolayer graphene is shown in Fig.~\ref{Fig2}(c). The incident angle is chosen as $\theta_i=56.6^\circ$ which is to improve the sensitivity to the optical conductivity~\cite{Chen2015,Liu2017}. The Rayleigh length in our case is $z_r\approx2.23\mathrm{mm}$ with the beam waist $21\mathrm{\mu{m}}$. For $\beta=2^\circ$, $3^\circ$, and $5^\circ$, the amplified factors $F\cot\beta\approx3200$, $2140$, and $1280$ are obtained with the distance of free evolution $z=250\mathrm{mm}$. Another important parameter (susceptibility $\chi$) in graphene can also cause the current density due to the change of polarization with time in atomic sheets. In Ref.~\cite{Merano2016}, Merano use the experimental data to fit susceptibility as $8\times10^{-10}\pm3\times10^{-10}\mathrm{m}$ in the visible spectral range~\cite{Kravets2010}. the amplified pointer shifts with the variation of $\chi$ from $6\times10^{-10}$ to $10\times10^{-10}\mathrm{m}$ in Fig.~\ref{Fig2}(d) are given. Therefore, the contribution of the susceptibility in the amplified pointer shifts can be neglected.

\begin{figure}
\includegraphics[width=8cm]{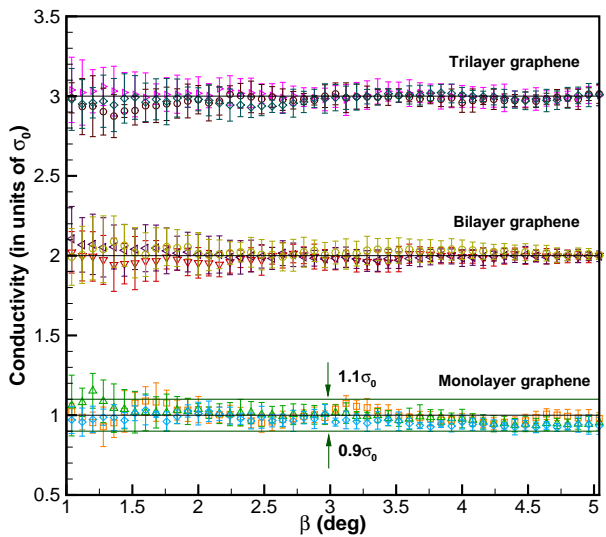}
\caption{\label{Fig5} Measurement of the optical conductivity in units of $\sigma_0$ for monolayer, bilayer, and trilayer graphene. Theoretically, the  optical conductivity of monolayer graphene is taken as the universal constant $\sigma_0=e^2/4\hbar$ shown as the bottom black horizontal line. The green horizontal lines correspond to the value of $0.9\sigma_0$ and $1.1\sigma_0$. In addition, the optical conductivities of bilayer and trilayer graphene without twist are also obtained.}
\end{figure}

In our experiment, we measure the conductivity of graphene samples supported on bulk $SiO_2$ substrates. The area of the region containing graphene on the substrate is about $1\mathrm{cm}\times1\mathrm{cm}$. The experiment setup to observe the photonic SHE via weak measurements is shown in Fig.~\ref{Fig3}. A Gaussian beam with wavelength of $633\mathrm{nm}$ is reflected by a graphene-$SiO_2$ interface to produce the photonic SHE. The the beam waist of light beam after L1 is $21\mu\mathrm{m}$. The spin displacements are much smaller than the width of beam, so the meter state of different spin eigenstates almost overlap to each other. This provide the weak interaction in weak measurements, which is associated with the optical conductivity. The polarization states is selected by the Glan laser polarizers. We use GLP1 to preselect the incidence state $|\psi_{i}\rangle$, and the combination of the quarter-wave plate (QWP) and GLP2 is to postselect the final state $|\psi_{f}\rangle$ to obtain a complex weak value. The rotations of the optical elements corresponding to the projection of the states on the Poincar\'{e} sphere are shown in the insets. In our system the purely imaginary weak value can result in a larger amplified factor due to effective light propagation. Hence, QWP for obtaining the real weak value dose not require, indicating $\alpha=0$. After the postselection of GLP2, the meter distribution with large displacement can be observed by a CCD camera.

The amplified pointer shift is measured with the rotation angle of GLP2 from $0^\circ$ to $5^\circ$. We investigate each graphene sample with three different incidence position. The three groups of data of monolayer graphene with error bar are represented by dots in Fig.~\ref{Fig4}.  We obtained the corresponding fitting curves to characterize the optical conductivity from the experimental results. In Ref~\cite{Merano2016}, the ensemble of the experimental data extracted from some remarkable experiments can fit a optical conductivity $(1.00\pm0.33)\sigma_0$ for the spectral range $450\sim750\mathrm{nm}$. And a optical conductivity $(1.0\pm0.1)\sigma_0$ was obtained by observing the frequency-independent absorbance of graphene in Ref~\cite{Mak2008}. To show the measurement accuracy of our method, the theoretical curves of amplified shifts for the optical conductivity $0.67\sigma_0$, $0.9\sigma_0$, $1.1\sigma_0$, and $1.33\sigma_0$ are given. The CCD we use has a displacement resolution of $1\mu\mathrm{m}$. Based on Eq.~(\ref{AS}) with $\beta=2^{\circ}$, it can lead to the minimum resolution of optical conductivity as $1.5\times10^{-8}\Omega^{-1}$. By fitting curves extracted from weak measurements, a high resolution about $\pm0.005\sigma_0$ in our method can be achieved. The amplified pointer shifts of bilayer and trilayer graphene are also measured, and the experimental results are provided in the Supplemental Material.

From the well-established theoretical model, the real part of the optical conductivity of monolayer graphene is a steplike function of frequency at zero temperature, and there is no imaginary  part in high-frequency region~\cite{Mikhailov2007}. The environment such as temperature and chemical potential only affect conductivity near the steplike point. For monolayer graphene, the bottom black horizontal line in Fig.~\ref{Fig5} indicates the conductivity have a universal constant value. The green horizontal lines stands for the value $0.9\sigma_0$ and $1.1\sigma_0$ to highlight our measuring accuracy. It is shown that the deviation of experimental results become smaller when the measured angle $\beta$ is large enough. This is due to the feature of the amplified shifts that are indistinguishable at small $\beta$, as shown in Fig.~\ref{Fig4}. For few-layer graphene without twist, the results in Fig.~\ref{Fig5} show that the conductivities of bilayer and trilayer graphene are close to the values of $2\sigma_0$ and $3\sigma_0$, respectively. In addition to the experiment of measuring the optical conductivity described above, Raman spectra with the Raman peak to confirm the region of graphene is measured (see the Supplemental Material).

In conclusion, the optical conductivity of graphene has been precisely measured based on the photonic SHE incorporating with the weak-value amplification technique. In our method, the spin-dependent shifts in photonic SHE serve as a measurement pointer, and the weak-value amplification technique is to obtain an amplified pointer shift. The experimental results show that the optical conductivity of monolayer graphene is close to the universal constant value $(0.993\pm0.005)\sigma_0$ with a high measuring resolution $1.5\times10^{-8}\Omega^{-1}$. For bilayer and trilayer graphene, the twist between the atomic sheets is not considered in our case. We have found that the optical conductivity of few-layer graphene increasing linearly with layer number. It would be a very interesting idea to measure the optical conductivity of chiral graphene~\cite{Kim2016} and magic-angle graphene~\cite{Cao2018,Kamp2018}.

\section*{ACKNOWLEDGMENTS}
This work was supported by the National Natural Science Foundation of China (Grant No. 61835004); Fundamental Research Funds for the Central Universities (Grant No. 531118010313).

\end{document}